\documentclass[a4paper,10pt,twocolumn,showpacs,preprintnumbers]{revtex4}
\usepackage{graphicx,amsmath,amssymb}

\begin{document}

\title{Superstatistical fluctuations in time series: Applications to 
share-price dynamics and turbulence}

\author{Erik Van der Straeten}
\author{Christian Beck}
\affiliation{Queen Mary University of London, School of Mathematical Sciences, Mile End Road, London E1 4NS, UK}
\email[E-mail:\ ]{e.straeten@qmul.ac.uk, c.beck@qmul.ac.uk}

\date{\today}

\begin{abstract}
We report a general technique to study a given experimental 
time series with superstatistics. Crucial for the applicability of 
the superstatistics concept is the existence of a parameter $\beta$ that fluctuates on a large time scale as compared to the other time scales of the complex system under consideration. The proposed method extracts the main superstatistical parameters out of a given data set and examines the validity of the superstatistical model assumptions. We test the method thoroughly with surrogate data sets. Then the applicability of the superstatistical approach is illustrated using real experimental data. We study two examples, velocity time series measured in turbulent Taylor-Couette flows and time series of log returns of the closing prices of some stock market indices. 
\end{abstract}


\maketitle

\section{Introduction}
Superstatistical techniques as introduced in \cite{ref01} can be applied to large classes of nonequilibrium systems. The crucial assumption of this approach is that the dynamics of the system under study is a superposition of two dynamics with well separated time scales. The standard example is a Brownian particle moving through a slowly fluctuating environment (e.g. there is a slowly fluctuating temperature $1/\beta$). The fast dynamics is then determined by the change of the velocity of the Brownian particle and the slow dynamics is determined by the change of the temperature of the environment. It is well known that the stationary velocity distribution of a Brownian particle in a constant environment is a Gaussian distribution with variance $1/\beta$. The theory of superstatistics applies when the slow dynamics is so slow that the velocity distribution of the particle has time to relax to a Gaussian distribution between the changes of the environment. As such, after a long time, the stationary velocity distribution of the particle is just a superposition of Gaussian distributions weighted with a function $f(\beta)$. This $f(\beta)$ is the probability density to observe some value of $\beta$. Depending on $f(\beta)$, different results for the stationary velocity distribution will occur. An important question is which types of distributions will occur in 'typical' complex systems. In \cite{ref02}, the authors give some probabilistic arguments in favor of three distributions, the lognormal distribution, the gamma distribution and the inverse gamma distribution. One can also derive these distributions from a maximum entropy principle \cite{ref03}. The present paper will focus on techniques to extract the distribution $f(\beta)$ and the relevant superstatistical parameters out of real experimental data. 

Recent applications of superstatistical methods include hydrodynamic turbulence \cite{ref04,ref05,ref06,ref02}, pattern forming systems \cite{ref07}, cosmic rays \cite{ref08}, solar flares \cite{ref09}, share-price fluctuations \cite{ref10,ref11,ref12,ref13}, random matrix theory \cite{ref14,ref15}, random networks \cite{ref16}, multiplicative-noise stochastic processes \cite{ref17}, wind velocity fluctuations \cite{ref18}, hydro-climatic fluctuations \cite{ref19}, the statistics of waiting times \cite{ref20a,ref20} and models of the metastatic cascade in cancerous systems \cite{ref21}. For the present paper, we will concentrate on time series as generated by hydrodynamic turbulence and share-price fluctuations. 

The outline of the paper is as follows. In the next section we fix our notation. In section \ref{section_log_ori} we discuss the techniques proposed in \cite{ref02} to extract superstatistical parameters out of a given time series. We propose a modification to this method in section \ref{mod_zelf}. In section \ref{section_ex} we test our method using surrogate data sets and show that our modification improves
on the previously proposed method \cite{ref02}. In sections \ref{section_dist_fb} and \ref{sect_eco} we apply the superstatistical approach to real experimental data from hydrodynamic turbulence and finance. An alternative procedure to study superstatistical time series is briefly examined in section \ref{alt_rul}. The final section contains a discussion of our results.

\section{Notation and superstatistical approach}\label{notation}
The starting point is a given discrete time series $u$ containing $n$ data points. The different data points are denoted as $u_i$ with $i=1,2,\ldots n$. The probability distribution of the random variable $u$, extracted from the experimental data, is denoted as $P(u)$. The total time series is divided into $N$ equal slices of length $\Delta$, with $N=\lfloor n/\Delta\rfloor$ where $\lfloor x\rfloor$ means rounding the value of $x$ to the nearest lower integer. This implies that the $l$th time slice contains the measurement points $u_i$ with $1+(l-1)\Delta\leq i\leq l\Delta$. We then define 'local' moments of order $k$ as follows
\begin{eqnarray}
\langle u^k\rangle_{\Delta,l}=\frac1{\Delta}\sum_{j=1+(l-1)\Delta}^{l\Delta}u_j^k, &\textrm{with}&l=1,2,\ldots N.
\end{eqnarray}
In the following, we will assume that the first momentum of the total time series $u$
\begin{eqnarray}\label{ave_time}
\langle u\rangle_{n,1}&=&\frac1n\sum_{i=1}^nu_i
\end{eqnarray}
is equal to $0$. If this is not the case we use the time series $u-\langle u\rangle_{n,1}$ instead of $u$ itself.

Now we want to analyse the distribution $P(u)$ using superstatistics. We assume that the time series $u$ contains two different time scales $\tau$ and $T$ such that $\tau/T<<1$. Remember that we divided the time series $u$ in different time slices of length $\Delta$. The time scale $\tau$ determines how fast 'local' equilibrium is reached in these different slices. The time scale $T$ is the length of the slices. For every time slice, one can extract a 'local'  probability distribution for the variables $u_i$ with $1+(l-1)\Delta\leq i\leq l\Delta$ and $\Delta=T$. When $\tau/T<<1$, it is a good approximation to assume that 'local' equilibrium is reached in a slice of length $T$. By this we mean that the 'local' distributions can be approximated by Gaussian distributions
\begin{eqnarray}
p_{T,l}(u)&=&\sqrt\frac{\beta_{T,l}}{2\pi}e^{-\frac12\beta_{T,l} u^2},
\end{eqnarray}
with $\beta_{T,l}=1/\langle u^2\rangle_{T,l}$. Within this assumption, the distribution $P(u)$ is approximated by
\begin{eqnarray}\label{Hu_1}
P(u)\approx p_T(u):=\frac1N\sum_{l=1}^Np_{T,l}(u),
\end{eqnarray}
with $N=\lfloor n/T\rfloor$. In this way, one obtains $N$ values for $\beta_{T,l}$. When $N$ is large enough, one can replace expression (\ref{Hu_1}) by
\begin{eqnarray}\label{Hu_2}
P(u)\approx p_T(u)\approx p_{T,f}(u):=\int_0^\infty d\beta f_T(\beta)\sqrt\frac{\beta}{2\pi}e^{-\frac12\beta u^2},
\nonumber\\
&&
\end{eqnarray}
with $f_T(\beta)$ being the probability density that the value of the inverse variance in a randomly chosen time slice of length $T$ equals $\beta$.  The obtained distribution $f_T(\beta)$ depends on the long time scale $T$, because $T$ determines the length of the time slices and as such the obtained values for the parameter $\beta$. Therefore, the proper definition of $T$ is of crucial importance and a major part of the present paper will focus on this issue.

In \cite{ref02}, a method was introduced to extract $f_T(\beta)$ out of a given experimental time series. The authors proposed to check the validity of their approach a posteriori by comparing the resulting distribution $p_{T,f}(u)$, with the distribution $P(u)$ extracted from the experimental data. Notice that the superstatistical approach includes two approximations. In the first step, one assumes the existence of two time scales $\tau$ and $T$ such that in every time slice 'local' equilibrium is reached. In the second step one assumes the existence of a distribution $f_T(\beta)$ replacing the summation in expression (\ref{Hu_1}) by an integral. One would like to judge the validity of the first approximation before moving on to the second approximation. Therefore, we propose an extension of the method introduced in \cite{ref02} to examine the quality of the first approximation. Afterwards, the validity of the second approximation can be checked by comparing the distributions $p_T(u)$ and $p_{T,f}(u)$. In the present paper we investigate the two steps of the superstatistical approach separately. We will test our arguments with surrogate data and with real experimental data sets.

\section{Original definition of time scales}\label{section_log_ori}
The correlation function of a time series $u$ can be calculated as follows
\begin{eqnarray}
C_{n,t}(u)&=&\frac1{n-t}\sum_{i=1}^{n-t}u_iu_{i+t}.
\end{eqnarray}
The superstatistical short time scale $\tau$ of the time series is defined by the exponential decay of $C_{n,t}(u)$ \cite{ref02}
\begin{eqnarray}\label{kortetijd}
C_{n,\tau}(u)&=&e^{-1}C_{n,0}(u).
\end{eqnarray}
In \cite{ref02} a function $\kappa_{\Delta}$ is introduced as
\begin{eqnarray}\label{kazondergem}
\kappa_\Delta=\frac1N\sum_{l=1}^N\kappa_{\Delta,l},&\textrm{with}&\kappa_{\Delta,l}=\frac{\left\langle u^4\right\rangle_{\Delta,l}}{\left\langle u^2\right\rangle_{\Delta,l}^2}.
\end{eqnarray}
Notice that $\kappa_{\Delta,l}$ is just the kurtosis of the $l$th time slice. The superstatistical long time scale $T$ is then defined by the condition
\begin{eqnarray}\label{def_lon}
\kappa_T&=&3.
\end{eqnarray}
To understand the meaning of this definition, remember that the main assumption of superstatistics is the existence of two well separated time scales. When this is true, one can approximate the distribution of the variables in the $l$th time slice by a Gaussian distribution (in this paper we always assume
that 'local' equilibrium is associated with
Gaussian behavior). When the variables are indeed locally
Gaussian distributed with zero mean and variance $1/\beta_{\Delta,l}$, the first four 'local' moments and $\kappa_{\Delta,l}$ are simply
\begin{eqnarray}\label{moments}
&&\langle u\rangle_{\Delta,l}=0,\ \langle u^2\rangle_{\Delta,l}=\frac1{\beta_{\Delta,l}},\ \langle u^3\rangle_{\Delta,l}=0,
\cr
&&\langle u^4\rangle_{\Delta,l}=\frac3{\beta_{\Delta,l}^2},\ \kappa_{\Delta,l}=3.
\end{eqnarray}
The condition (\ref{def_lon}) implies that one is looking for a suitable division of the total data set into time slices for which the variables are locally Gaussian distributed \cite{ref02}, with a variance that fluctuates
from slice to slice.

This definition will always result in a value for the long time scale whenever the kurtosis of the complete time series $\kappa_{n,1}$ is larger than $3$ \cite{ref02,ref22}. This can be seen by considering two special cases. In the first case, $\Delta$ is so small that only one value of $u$ is observed in each time slice. This results in $\kappa_1=\kappa_{1,l}=1$ (with $l=1,2,\ldots n$). In the second case, $\Delta$ is so large that it includes the entire time series $u$. This results in $\kappa_n=\kappa_{n,1}>3$. As a consequence, there exists a long time scale $1<T<n$ for which $\kappa_\Delta=3$ holds. With the definitions proposed in \cite{ref02}, one could formally use superstatistics to analyse the distribution of any time series whenever $\tau/T<<1$ and $\kappa_{n,1}>3$. But clearly not every time series that fulfills these conditions contains time slices of which one can assume that the variables are locally Gaussian distributed. Therefore, in the next section we will derive an extra condition that should hold before we assume that a time series at hand can be described by superstatistics.

\section{Extra constraint}\label{mod_zelf}
Assume we found the value of $\Delta$ for which $\kappa_\Delta$ equals $3$. This value is denoted as $T$. This means that the complete time series is divided in $N=\lfloor n/T\rfloor$ time slices of length $T$. The second and fourth momentum of these time slices are denoted as $\langle u^2\rangle_{T,l}$ and $\langle u^4\rangle_{T,l}$ respectively. Then one can express the value of the function $\kappa_\Delta$ for $\Delta=NT$ as follows
\begin{eqnarray}\label{kappa_ef}
\kappa_{NT}&=&\left(\frac1N\sum_{j=1}^N\langle u^2\rangle_{T,j}\right)^{-2}\frac1N\sum_{j=1}^N\langle u^4\rangle_{T,j}.
\end{eqnarray}
For large data series, the kurtosis of the complete data set $\kappa_{n,1}$ is in good approximation equal to $\kappa_{NT}$ because $N=\lfloor n/T\rfloor\approx n/T$. Formula (\ref{kappa_ef}) shows that the value of $\kappa_{n,1}$ can be calculated starting from the second and fourth momentum of the time slices of length $T$. Now we assume that these time slices contain variables that are approximately Gaussian distributed. We will determine the influence of the deviations from this approximation to the value of $\kappa_{n,1}$. Remember that the moments of purely Gaussian distributed random variables are given by (\ref{moments}). However, the variables of a time series are never perfectly Gaussian distributed. We define $\theta_{T,l}$ as the deviation of the fourth momentum from $3\langle u^2\rangle_{T,l}^2$ for the time slices of length $T$
\begin{eqnarray}\label{def_theta}
\theta_{T,l}=\langle u^4\rangle_{T,l}-3\langle u^2\rangle_{T,l}^2.
\end{eqnarray}
Then expression (\ref{kappa_ef}) can be rewritten as follows
\begin{eqnarray}
\kappa_{NT}&=&\left(\frac1N\sum_{j=1}^N\langle u^2\rangle_{T,j}\right)^{-2}\frac1N\sum_{j=1}^N\left[3\langle u^2\rangle_{T,j}^2+\theta_{T,j}\right].
\nonumber\\
&&
\end{eqnarray}
When the Gaussian approximation is reasonable in the time slices of length $T$, the contribution of the term $\sum_j\theta_{T,j}$ to the value of $\kappa_{NT}$ will be small as compared to the contribution of the term $3\sum_j\langle u^2\rangle_{T,j}^2$. Therefore, we expect the Gaussian approximation to hold when $|\epsilon|<<1$ with
\begin{eqnarray}\label{rule_extra}
\epsilon&=&\frac13\frac{\sum_{j=1}^N\theta_{T,j}}{\sum_{j=1}^N\langle u^2\rangle_{T,j}^2}.
\end{eqnarray}
We showed that the value of the kurtosis of the complete data set $\kappa_{n,1}$ can be expressed as a function of the moments of the times slices of length $T$ only. The parameter $\epsilon$ basically measures the contribution of the deviations from the Gaussian approximation in these time slices to the value of $\kappa_{n,1}$. More details about the derivation and the interpretation of expression (\ref{rule_extra}) can be found in appendix \ref{appa}. In the next section we will show that the extra constraint $|\epsilon|<<1$ resolves some of the ambiguities of the original approach \cite{ref02}.

\section{Surrogate data}\label{section_ex}
The major problem of the definition (\ref{def_lon}) of the long time scale $T$ is that it will always give a value for $T$ whenever the kurtosis of the complete time series $\kappa_{n,1}$ is larger than $3$. For example, a time series containing Gaussian distributed random variables 
of constant variance with just one outliner can have $\kappa_{n,1}>3$ and $\tau/T<<1$ without being superstatistical. Also a time series containing q-Gaussian 
\cite{ref23} distributed random variables with constant variance can be wrongly classified as being superstatistical. In this section we use surrogate time series to illustrate that the extra constraint $|\epsilon|<<1$ is a useful tool to decide whether a given time series is superstatistical. We study three examples, (A) data constructed by numerically integrating a discretized Langevin equation, (B) time series containing outliners and (C) data sets containing q-Gaussian distributed random variables.

\subsection{Langevin-like surrogate data}\label{langee}
\begin{figure}
\begin{center}
\parbox{0.48\textwidth}{\includegraphics[width=0.48\textwidth]{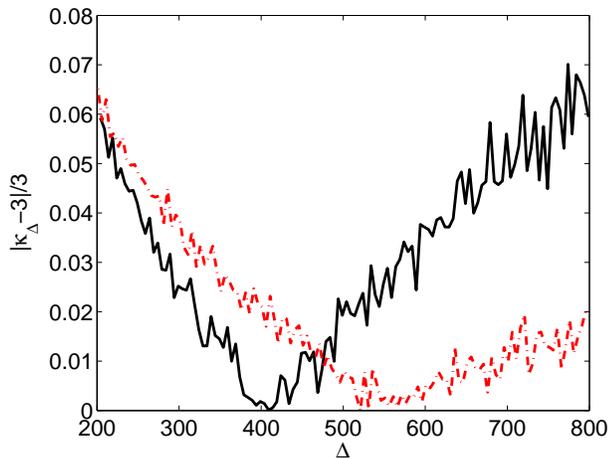}}
\caption{\label{fig01} (Colour online) Extraction of the long time scale $T$
from the condition $\kappa_\Delta=3$ for two surrogate data sets (see expression (\ref{kazondergem}) for the definition of $\kappa_\Delta$). The difference between the two time series is the width of the distribution of the fluctuating parameter $\beta$, see Fig.~\ref{fig02}.}
\end{center}
\end{figure}
\begin{figure}
\begin{center}
\parbox{0.48\textwidth}{\includegraphics[width=0.48\textwidth]{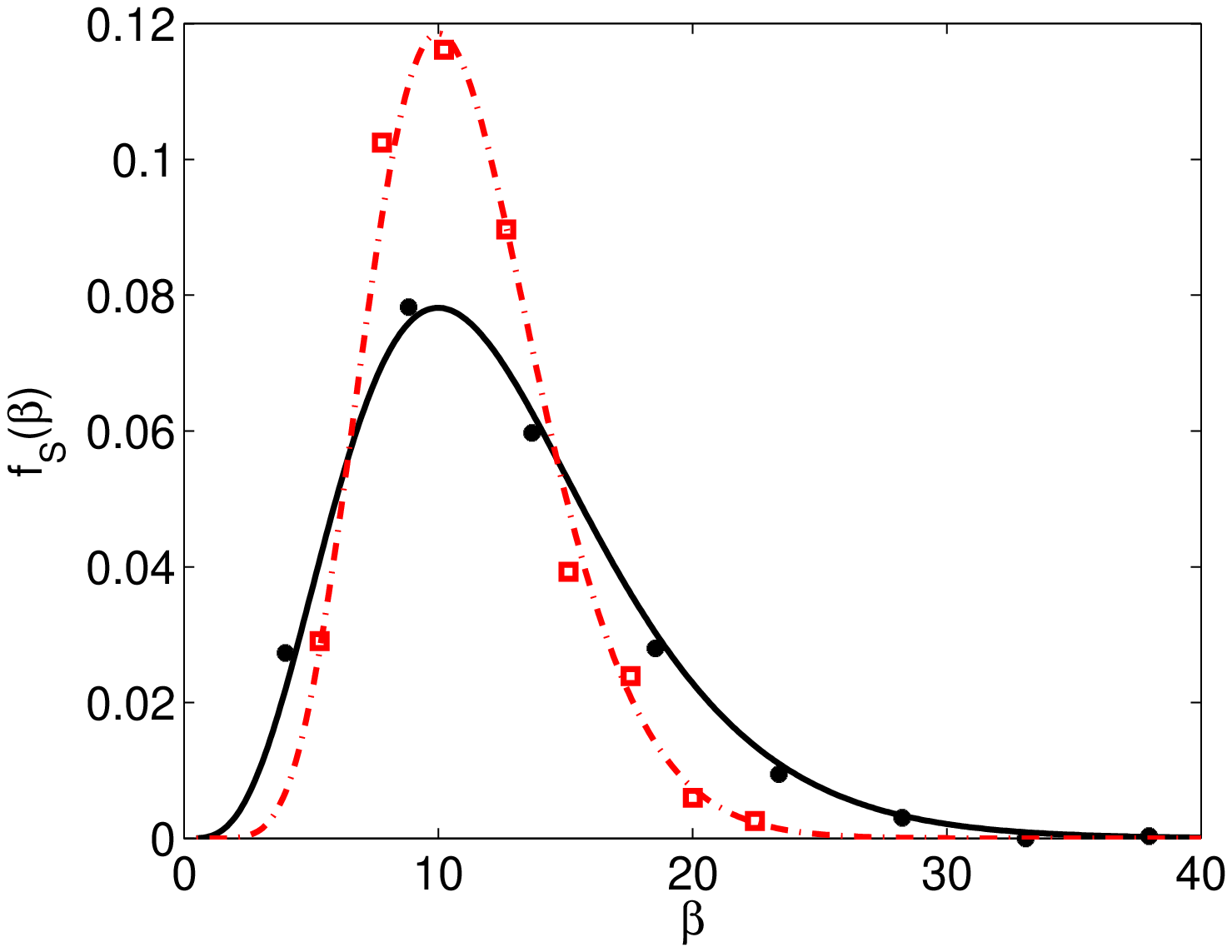}}
\hfill
\parbox{0.48\textwidth}{\includegraphics[width=0.48\textwidth]{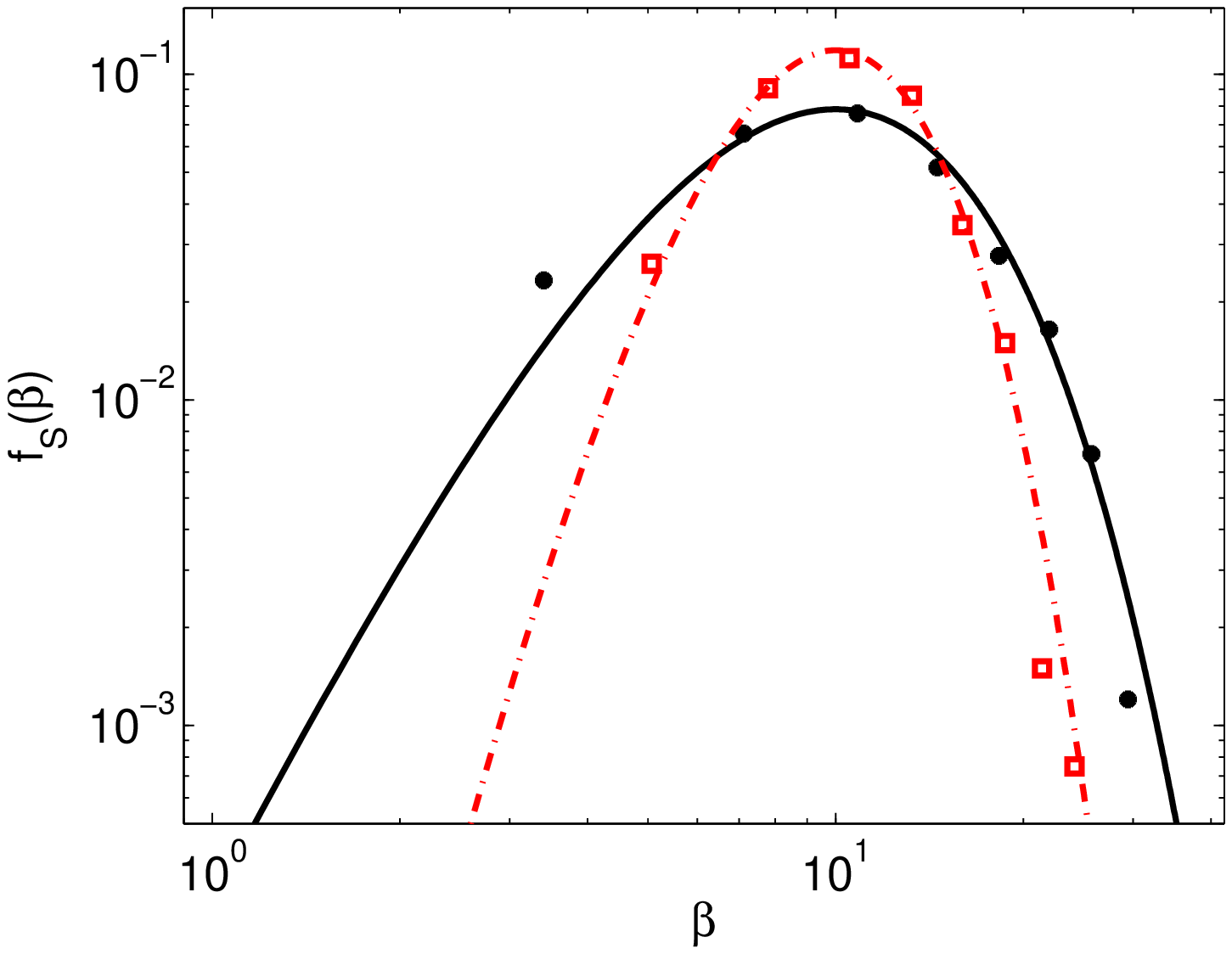}}
\caption{\label{fig02} (Colour online) The black solid line and the red dashed-dotted line are representations of the distribution (\ref{fbeta_surro12}) with $\theta=10/(\alpha-1)$ and $\alpha=5,10$ respectively. These distributions are used to construct two surrogate time series. Applying the superstatistical approach to these data sets results in an approximation $f_T(\beta)$ for the imposed distribution $f_S(\beta)$. The extracted distributions $f_T(\beta)$ are represented by black dots and red squares for the data sets with $\alpha=5,10$ respectively.}
\end{center}
\end{figure}
\begin{figure}
\begin{center}
\parbox{0.48\textwidth}{\includegraphics[width=0.48\textwidth]{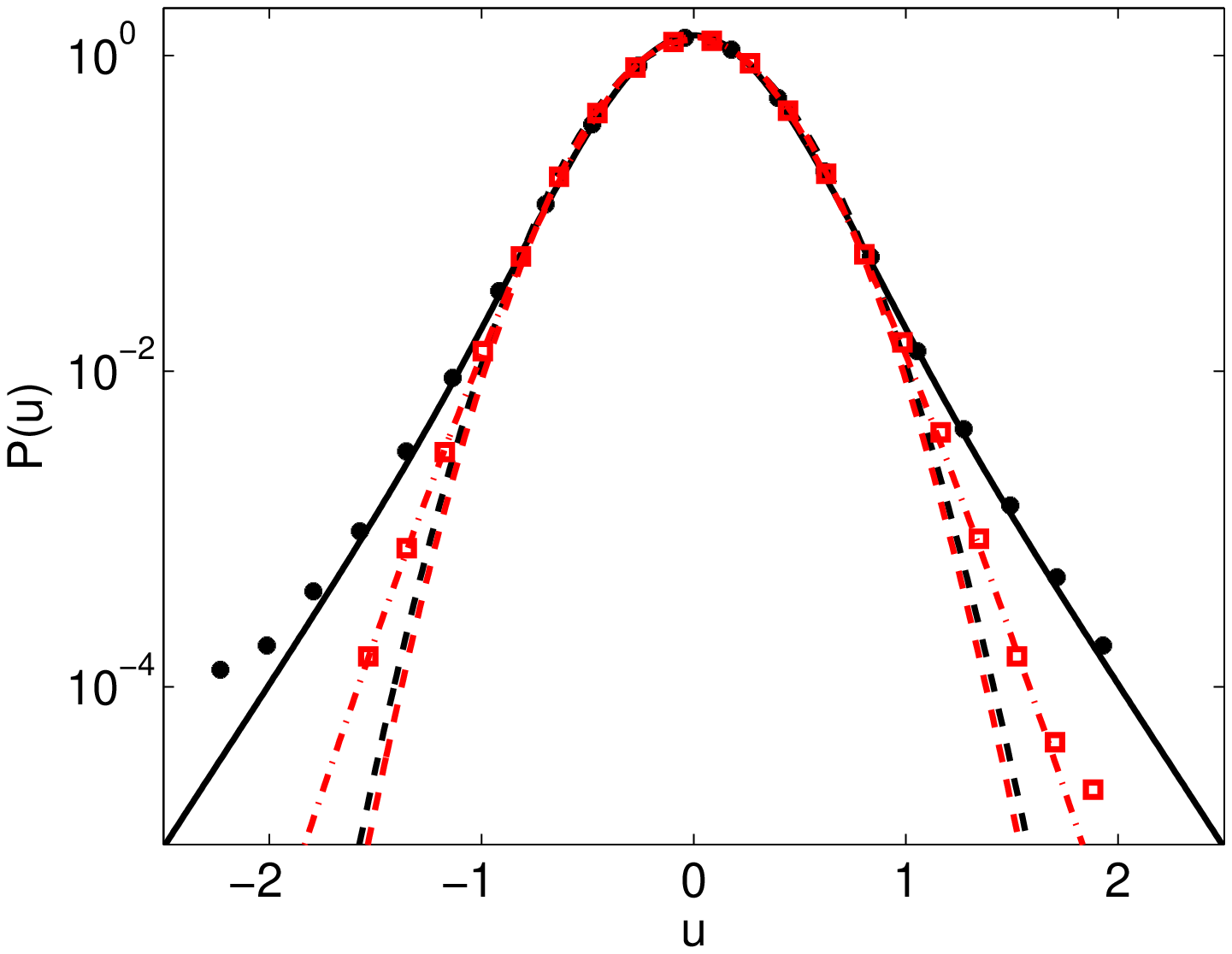}}
\caption{\label{fig03} (Colour online) The black dots and the red squares represent the observed distributions $P(u)$ for the surrogate data sets with $\alpha=5,10$ respectively. The superstatistical approximations $p_T(u)$ are given by the black solid line and the red dashed-dotted line. The dashed lines are Gaussians with the same variance as $P(u)$.}
\end{center}
\end{figure}
In order to construct realistic surrogate data sets, we numerically iterate a discretized Langevin equation
\begin{eqnarray}\label{langevin_dis}
u_i=u_{i-1}-\gamma u_{i-1}+\sqrt{\frac{2\gamma}{\beta_i}}L_i,
\end{eqnarray}
where $\gamma$ is a constant, the $L_i$ correspond to Gaussian white noise with variance $1$ and $\beta_i$ are random variables with a certain distribution. This is the dynamics of the example discussed in the introduction, a Brownian particle moving through a slowly fluctuating environment. When $\beta_i=\beta$ is chosen to be a constant, one ends up with the standard Langevin equation. In this special case, it is known that the stationary probability density of $u$ is a Gaussian distribution with inverse variance $\beta$. The relaxation time is $1/\gamma$. We study the more general case in which $\beta_i$ is a random variable that is gamma distributed
\begin{eqnarray}\label{fbeta_surro12}
f_S(\beta)&=&\frac{\theta^{-\alpha}}{\Gamma(\alpha)}\beta^{\alpha-1}e^{-\beta/\theta},
\end{eqnarray}
with $\alpha$ and $\theta$ constants. However, in order to make the resulting surrogate time series superstatistical, the time scale $T_S$ of the fluctuating parameter $\beta_i$ must be larger than the time scale $\tau_S=1/\gamma$ on which 'local' equilibrium is reached. In this way, the probability distribution of $u$ will relax to a Gaussian distribution before the next change of the value of $\beta_i$ occurs. Therefore, we chose $\tau_S=10$ and $T_S=500$. In practice we keep the value of $\beta_i$ constant and iterate the Langevin equation over $T_S=500$ steps. Then we change the value of $\beta_i$ and iterate the discrete Langevin equation again over $T_S=500$ steps. We repeat this procedure $N_S=500$ times. The result of this procedure is a time series with $n_S=T_SN_S=250.000$ data points. We construct two data sets, with different values of the parameters of the distribution (\ref{fbeta_surro12}), $\alpha=5,10$ and $\theta=10/(\alpha-1)$. Then we apply the superstatistical approach to these time series. 

The resulting short time scales are $\tau=9.52,9.21$ for the values of the parameter $\alpha=5,10$ respectively. Fig.~\ref{fig01} shows $\kappa_\Delta$ as a function of $\Delta$ for the two surrogate data sets. The black solid line and the red dashed-dotted line are obtained for $\alpha=5,10$ respectively. The constraint $\kappa_\Delta=3$ results in the following values of the long time scale: $T=409,521$ for $\alpha=5,10$. The corresponding values of $|\epsilon|$ are equal to $0.014$ and $0.004$ respectively. Because both inequalities $\tau/T<<1$ and $|\epsilon|<<1$ hold, we conclude that the data series at hand can be described within the superstatistical approach, and that our method to extract the relevant time scales works very well.

We continue by extracting $f_T(\beta)$ out of the time series and compare the obtained distribution $f_T(\beta)$ with the imposed distribution $f_S(\beta)$. The result of this calculation can be seen in Fig.~\ref{fig02}. We conclude that there is an excellent agreement between $f_T(\beta)$ and $f_S(\beta)$ for both surrogate data sets. Finally, we construct the distribution $p_T(u)$ and compare the obtained distribution $p_T(u)$ with the original distribution $P(u)$, see Fig.~\ref{fig03}. There is an excellent agreement between $p_T(u)$ and $P(u)$ for both surrogate data sets.

\subsection{Outliners}\label{outll}
\begin{figure}
\begin{center}
\parbox{0.48\textwidth}{\includegraphics[width=0.48\textwidth]{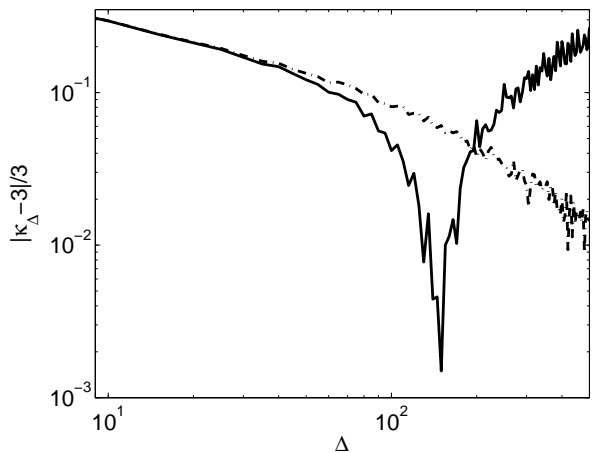}}
\caption{\label{fig04} Determination of the long time scale $T$ from the
condition $\kappa_\Delta=3$ for two surrogate data sets (see expression (\ref{kazondergem}) for the definition of $\kappa_\Delta$). The minimum of the solid line is reached for $\Delta=150$. The dashed line is a monotonic decreasing function of $\Delta$.}
\end{center}
\end{figure}
To illustrate the influence of an outliner to the superstatistical approach, we construct two surrogate data series. The first time series contains $n_S=25.000$ Gaussian distributed random variables constructed by numerically iterating a discretized Langevin  (\ref{langevin_dis}) with $\gamma=1/5$ and $\beta_i=\beta=1$. Clearly, the extracted value of the kurtosis for this time series is approximately equal to $3$. The second surrogate data series is constructed out of the first by replacing one of the elements of the first time series by an outliner. As a consequence of this, the kurtosis of the second time series will generally be larger than $3$ and is equal to $4.63$ for our particular example. Therefore, expression (\ref{def_lon}) would formally give a value for the long time scale when applied to the second data series. This can be seen in Fig.~\ref{fig04} that shows a plot of $\kappa_\Delta$ as a function of $\Delta$, computed from the first (dashed line) and the second (solid line) surrogate data set. For the first data set, $|\kappa_\Delta-3|/3$ is a monotonic decreasing function of $\Delta$, while for the second data set the function $|\kappa_\Delta-3|/3$ reaches a single minimum at $T=\Delta=150$. We also computed the short time scale for this surrogate data set, $\tau=4.52$. As a consequence, the inequality $\tau/T<<1$ holds. Finally we calculated the value of $|\epsilon|$ for the second surrogate data series and obtained $0.41$. Clearly, the inequality $|\epsilon|<<1$ does not hold in this case and the data series at hand cannot be described with the superstatistical approach. After removing the outliner, one is left with a data series containing just Gaussian variables of constant variance (equivalent to the surrogate data set we started from) and no value for the long time scale can be found, see Fig.~\ref{fig04}. This example shows that one has to be careful when using the superstatistical approach to study data series that contain outliners. Apparently, our criterion $|\epsilon|<<1$ can help to identify truly superstatistical dynamics.

\subsection{q-Gaussian}\label{subsection_ts}
\begin{figure}
\begin{center}
\parbox{0.48\textwidth}{\includegraphics[width=0.48\textwidth]{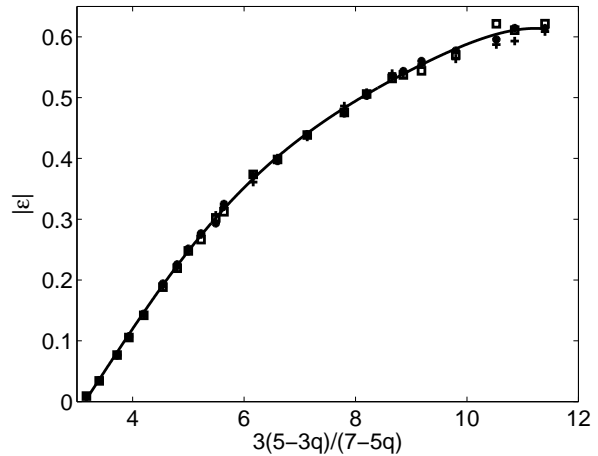}}
\caption{\label{fig05} Plot of the obtained value of the superstatistical parameter $|\epsilon|$ as a function of the kurtosis of the used q-Gaussian
surrogate data set. The solid line is a guide to the eye.}
\end{center}
\end{figure}
We continue by constructing surrogate data sets containing 
q-Gaussian random variables of constant variance. This means that the random variables are distributed according to the following distribution
\begin{eqnarray}\label{ts}
P(u)&=&\frac{\Gamma\left(\frac1{q-1}\right)}{\Gamma\left(\frac1{q-1}-\frac12\right)}\sqrt\frac{(q-1)b}{2\pi}\times
\nonumber\\
&&\left(1+(q-1)\frac12 b u^2\right)^{-1/(q-1)},
\end{eqnarray}
with $b$ and $q$ constants. In the limit $q\rightarrow1$, the distribution (\ref{ts}) approaches a Gaussian distribution with inverse variance $b$.  The kurtosis of the distribution (\ref{ts}) is $3(5-3q)/(7-5q)$. Clearly, the kurtosis is larger than $3$ for $q>1$. Therefore, expression (\ref{def_lon}) would formally give a value for the long time scale when applied to data sets containing q-Gaussian random variables with $q>1$.
Our aim here is to identify this time series as being not
superstatistical, since there is no long time scale on which
the variance changes.

The kurtosis diverges in the limit $q\rightarrow7/5$. That is the reason why the analysis presented in this section will be restricted to data series containing q-Gaussian random variables with $1<q<7/5$. We use the generalized Box-M\"uller method, proposed in \cite{ref23}, to construct the surrogate data series. The number of elements of every data set is $n_S=25.000$. The used values of $q$ range from $1.05$ to $1.35$, while the used values of $b$ are $0.1,1,10$. We calculate the value of $\Delta$ for which the constraint $\kappa_\Delta=3$ holds and determine the corresponding value of $|\epsilon|$. For every tuple $(q,b)$, we repeated these calculations $500$ times (we constructed $500$ different data sets) and averaged the value of $|\epsilon|$ over these different runs. The result of these calculations can be seen in Fig.~\ref{fig05}. This figure shows a plot of the value of $|\epsilon|$ as a function of the kurtosis of the used q-Gaussian distribution. The figure contains three curves, for three different values of $b$,  $10$ ($\bullet\bullet\bullet$), $1$ ({\scriptsize$+++$}) and $0.1$ ({\tiny $\square\square\square$}). Clearly, the three curves are almost indistinguishable. More importantly, the curves are increasing functions of the kurtosis (determined
by $q$). This shows that the extra constraint $|\epsilon|<<1$ will reject the time series containing purely q-Gaussian distributed random variables as being superstatistical when the value of the kurtosis is large enough. When the value of the kurtosis is close to $3$, the value of $|\epsilon|$ will also
become small. However notice that in these cases the corresponding q-Gaussian distribution is also better and better approximated by the Gaussian distribution. This means that the assumption of 'local' equilibrium can be relaxed to 'global' equilibrium. As a consequence, there is no reason to the study these time series within the superstatistical approach, because the usual equilibrium statistical mechanics (i.e. Gaussians with constant variance) can be used in very good approximation.

\section{Turbulence data}\label{section_dist_fb}
\begin{figure}
\begin{center}
\parbox{0.48\textwidth}{\includegraphics[width=0.48\textwidth]{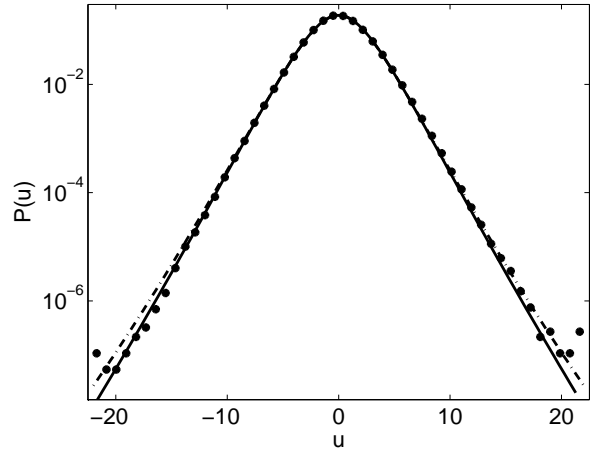}}
\caption{\label{fig06} Plot of the distribution $P(u)$ ($\bullet\bullet\bullet$) computed from a data series of velocity differences measured in turbulent Taylor-Couette flow.  The superstatistical approximations $p_T(u)$ and $p_{T,f}(u)$ are represented by the solid line and the dashed-dotted line respectively.}
\end{center}
\end{figure}
\begin{figure}
\begin{center}
\parbox{0.48\textwidth}{\includegraphics[width=0.48\textwidth]{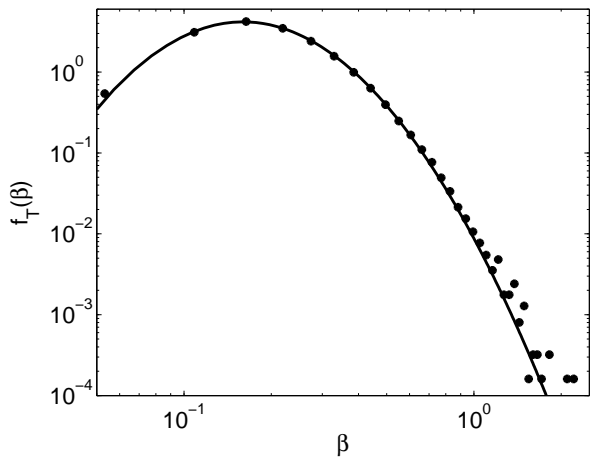}}
\caption{\label{fig07} Plot of the extracted distribution $f_T(\beta)$ ($\bullet\bullet\bullet$) computed from a data series containing the velocity differences measured in turbulent Taylor-Couette flow. The solid line shows a fit of $f_T(\beta)$ to a lognormal distribution, see expression (\ref{poi}), with $\alpha=0.5222$ and $\theta=-1.561$.}
\end{center}
\end{figure}
After testing our method with various types of surrogate data,
we now apply our superstatistical analysis
to real experimental data. We use time series obtained in an experiment 
performed by Lewis and Swinney \cite{ref24}. They measured a 
single velocity component $v(t)$ as a function of time $t$ in 
turbulent Taylor-Couette flow for different Reynolds numbers $Re$. 
The stationary probability distribution $P(u)$ of the velocity 
difference $u(t):=v(t+\delta)-v(t)$ at a given scale $\delta$ 
is well-known to exhibit non-Gaussian behavior. 

The turbulence data sets we used from Swinney's experiment are not
scale invariant. Contrary to other turbulence data there is no
inertial range for these data and no scaling exponents of
velocity increments can be defined. But in \cite{ref02} it was shown 
that the existence of two effective time scales is well
supported by the data.

The values of the parameters for the data series used in the present paper are $Re=540.000$, $\delta=16$ and $n\approx2\times10^7$. The two different time scales are 
extracted from the time series as
$\tau \approx 7.2$ and $T \approx 186$, in agreement with
the results of \cite{ref02}. We also evaluated expression (\ref{rule_extra}). The result is $|\epsilon|=0.0205<<1$. Hence we assume that the experimentally measured time series is superstatistical. 
We also constructed the distribution $p_T(u)$ which is an approximation for the real experimental distribution $P(u)$, see expression (\ref{Hu_1}). Fig.~\ref{fig06} shows the excellent agreement between these two distributions. 

As mentioned in section II, the superstatistical modeling approach includes two approximations. Until now, we only discussed the first. After this approximation one ends up with $N$ values of the 'local' inverse variance $\beta_{T,l}$. This series of numbers can be treated as a set of random variables with a certain distribution $f_T(\beta)$. In the second step of the superstatistical modeling process one tries to find the best fit to $f_T(\beta)$ with some well-known distributions such as, e.g. the lognormal distribution, the gamma distribution or the inverse gamma distribution,
thus proceeding to an analytic model. Fig.~\ref{fig07} shows 
our extracted distribution $f_T(\beta)$. A good fit
to the data is obtained for a lognormal distribution
\begin{eqnarray}\label{poi}
f_T(\beta)&=&\frac1{\alpha\sqrt{2\pi}}\frac1\beta e^{-(\ln\beta-\theta)^2/2\alpha^2},
\end{eqnarray}
with $\alpha=0.5222$ and $\theta=-1.561$ (see also \cite{ref03}).  
Finally we construct $p_{T,f}(u)$ (see expression (\ref{Hu_2})) and compare this distribution with $p_T(u)$. Fig.~\ref{fig06} includes plots of both $p_{T,f}(u)$ and $p_T(u)$. To summarize, Fig.~\ref{fig07} shows the good fit of a lognormal distribution to $f_T(\beta)$, while Fig.~\ref{fig06} shows the excellent agreement between $p_{T,f}(u)$ and $p_T(u)$. 
Both figures validate that lognormal superstatistics
is indeed a good modeling approach to the data.

\section{Economical data series}\label{sect_eco}
\begin{table}
\begin{tabular}{|c|cccc||cccc|}
\hline
&S\&P&500&&&DJI&&&
\\\hline
$\delta$&$\tau$&$T$&$|\epsilon|$&$\kappa_{1,n}$&$\tau$&$T$&$|\epsilon|$&$\kappa_{1,n}$
\\\hline
1&$<$1&17&1.11&35.6&$<$1&18&1.50&49.3
\\
2&1.28&33&1.12&22.6&1.27&32&1.28&28.0
\\
4&2.46&60&1.02&15.4&2.47&64&1.36&18.6
\\\hline
1&$<$1&17&0.02&6.6&$<$1&19&0.01&6.3
\\
2&1.29&32&0.01&6.1&1.29&34&0.01&6.0
\\
4&2.45&73&0.05&5.6&2.45&70&0.10&5.5
\\\hline
\end{tabular}
\caption{\label{tab} 
The values of the superstatistical parameters extracted from data series containing the normalized log returns $u_i$ of the DJI index and the S\&P 500 index for three different values of $\delta$. The used data sets are (top) unaltered $u_i$, (bottom) largest values of $|u_i|$ removed.}
\end{table}
\begin{figure}
\begin{center}
\parbox{0.48\textwidth}{\includegraphics[width=0.48\textwidth]{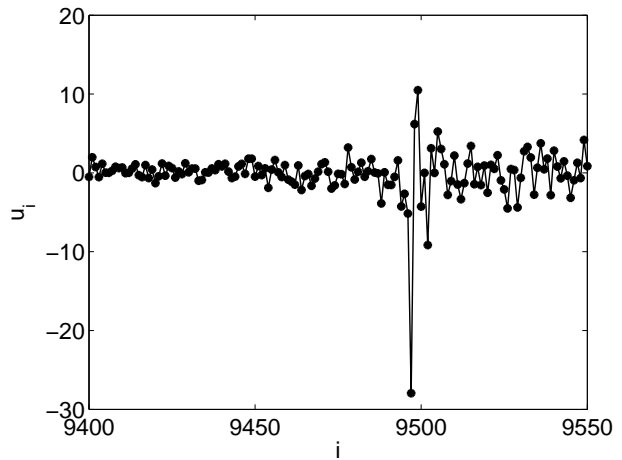}}
\caption{\label{fig08} Plot of the normalized log returns $u_i$ as a function of time $i$ for the DJI index with $\delta=1$. The plot shows just
a small section ($150$ data points) of the complete data series (approximately $15.000$ data points). Notice the outliner near $i=9500$ which corresponds to the crash of the stock markets on October 19, 1987.}
\end{center}
\end{figure}
\begin{figure}
\begin{center}
\parbox{0.48\textwidth}{\includegraphics[width=0.48\textwidth]{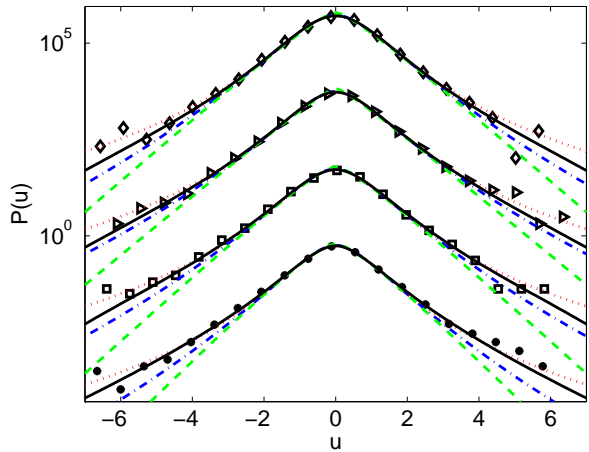}}
\caption{\label{fig09} (Colour online) Plot of the distributions $P(u)$ for the DJI index and the S\&P 500 index for two different values of $\delta$ [S\&P 500 index with $\delta=1$ ($\bullet\bullet\bullet$) and $\delta=2$ ({\tiny $\square\square\square$}), DJI index with $\delta=1$ ($\triangleright\triangleright\triangleright$) and $\delta=2$ ($\diamond\diamond\diamond$)].  The superstatistical approximations $p_T(u)$ are represented by the solid lines. The figure also shows the distributions $p_{T,f}$ for $3$ different choices of $f_T(\beta)$, the gamma distribution (dotted lines), the lognormal distribution (dashed dotted lines) and the inverse gamma distribution (dashed lines). The values of the
corresponding parameters $\alpha$ and $\theta$ are listed in Table~\ref{tab4}.}
\end{center}
\end{figure}
\begin{figure}
\begin{center}
\parbox{0.48\textwidth}{\includegraphics[width=0.48\textwidth]{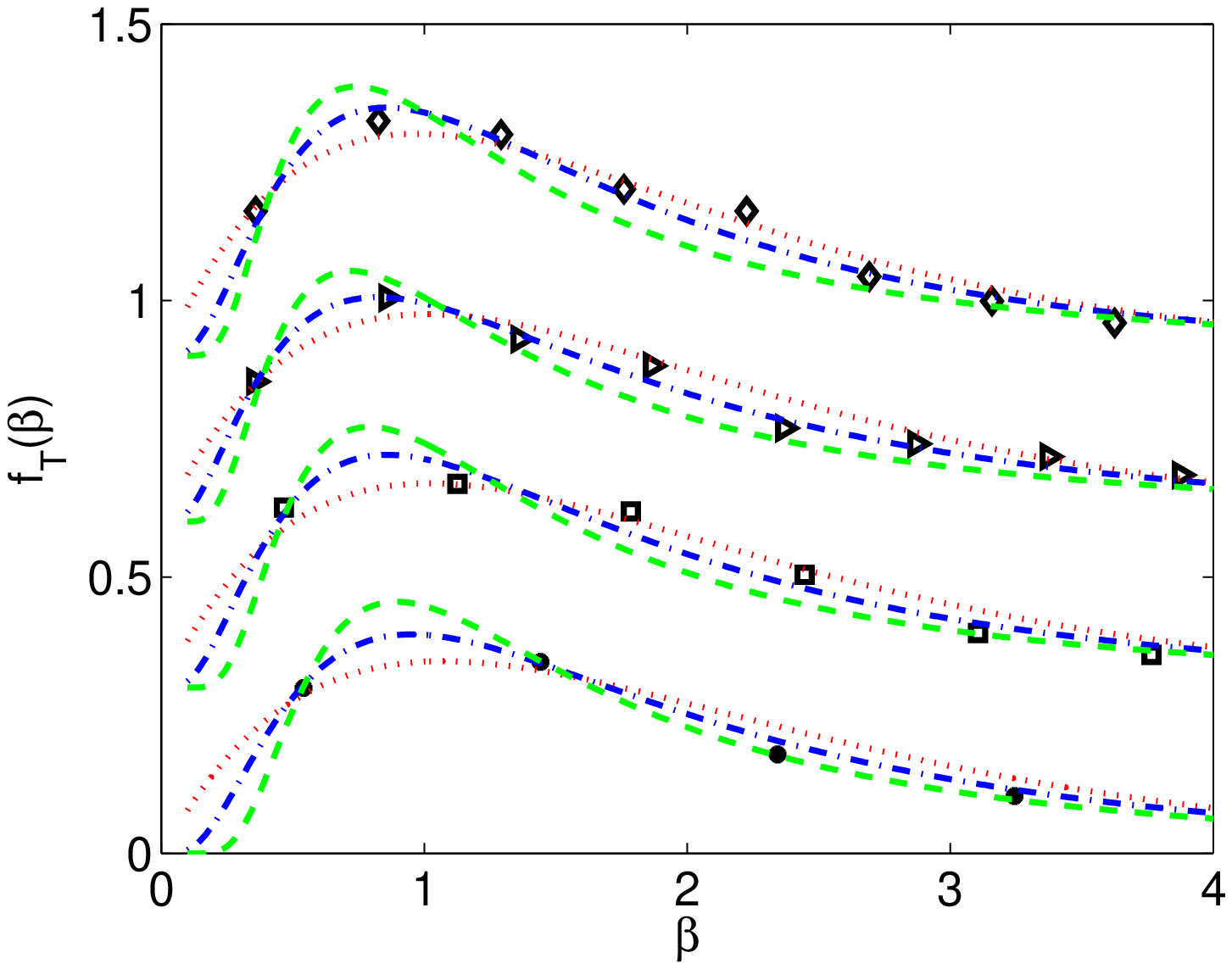}}
\hfill
\parbox{0.48\textwidth}{\includegraphics[width=0.48\textwidth]{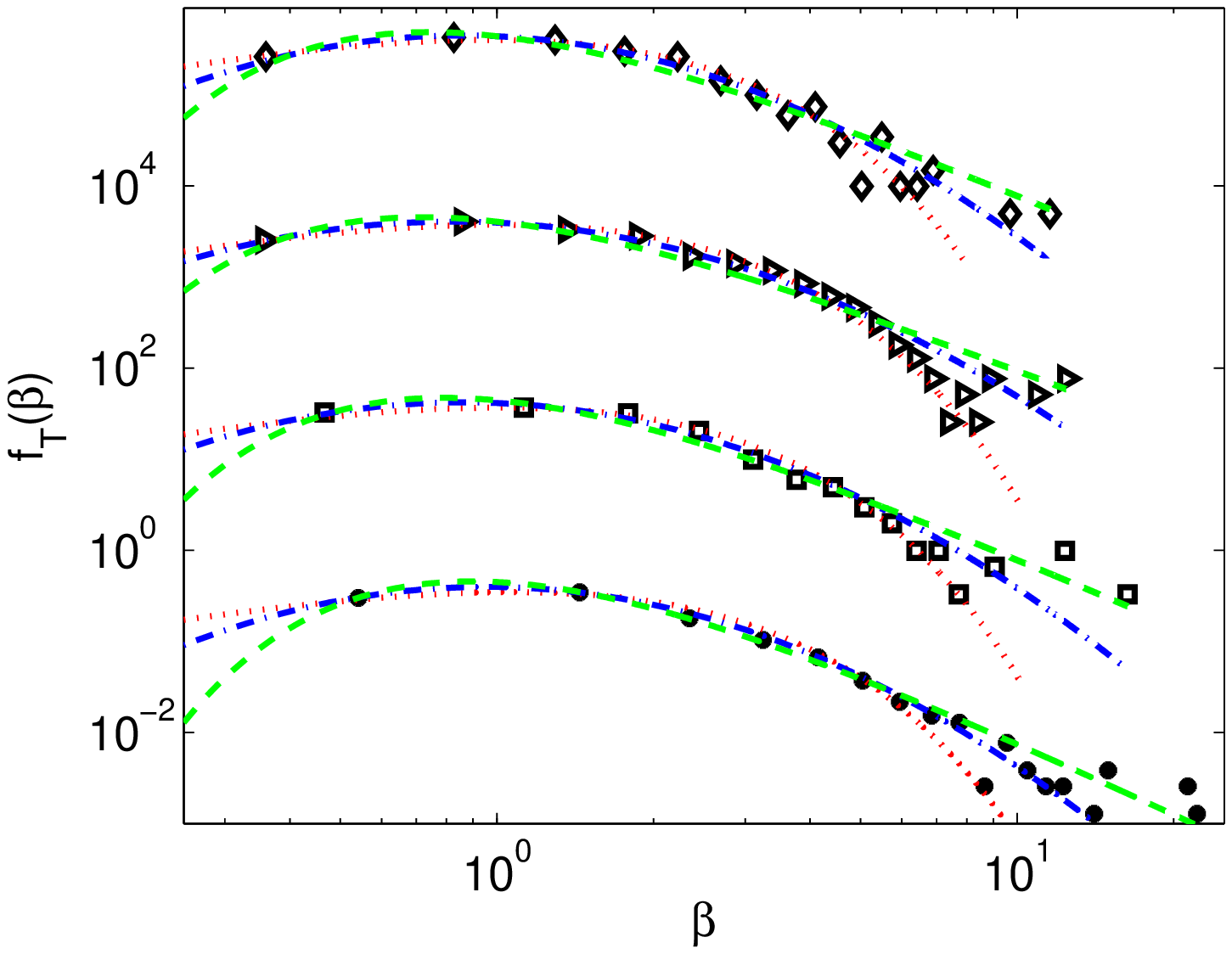}}
\caption{\label{fig10} (Colour online) Plot of the extracted distributions $f_T(\beta)$ for the DJI index and the S\&P 500 index for two different values of $\delta=1,2$ [S\&P 500 index with $\delta=1$ ($\bullet\bullet\bullet$) and $\delta=2$ ({\tiny $\square\square\square$}), DJI index with $\delta=1$ ($\triangleright\triangleright\triangleright$) and $\delta=2$ ($\diamond\diamond\diamond$)]. For all $4$ distributions an approximation to the data in terms of a gamma distribution (dotted lines), a lognormal distribution (dashed dotted lines) a inverse gamma distribution (dashed lines) is also shown. The values of the
corresponding parameters $\alpha$ and $\theta$ are listed in Table~\ref{tab4}.}
\end{center}
\end{figure}
In this section we study economical time series. The data sets are
the daily closing prices $x_i$ of the Dow Jones Industrial Average index (DJI) and the S\&P 500 index for the period March 1950 to September 2008. This means that the total number of data points is of the order $n\sim 15.000$. In the literature one usually studies the statistics of the log returns $X_i:=\ln(x_{i+\delta}/x_i)$ with $\delta=1,2,\ldots$ of the closing prices instead of the closing prices itself. We consider the normalized log returns
\begin{eqnarray}
u_i&:=&\left(X_i-\left\langle X\right\rangle\right)\left(\sqrt{\left\langle X^2\right\rangle-\left\langle X\right\rangle^2}\right)^{-1},
\end{eqnarray}
which have been rescaled to have variance $1$. An example of a part of such data series with $\delta=1$ for the DJI index can be seen in Fig.~\ref{fig08}. This figure illustrates that sometimes outliners (crashes of the stock markets) are present in the data. 

In the first step of the superstatistical analysis, we calculate $\tau$, $T$ and $|\epsilon|$. Additionally, we also calculate for every set the total kurtosis $\kappa_{1,n}$. We performed these calculations for the DJI index and the S\&P 500 index for three different values of $\delta=1,2,4$, see Table~\ref{tab}. For every value of $\delta$, the condition $\tau/T<<1$ holds, while the condition $|\epsilon|<<1$ does not hold. Notice also that the values of the total kurtosis strongly deviate from $3$. We illustrated in section \ref{outll} that outliners can strongly
influence the results of the superstatistical analysis. Therefore, we 
continue by studying data sets in which the data points with the largest values of $|u_i|$ are removed. Indeed, it
is well-known in mathematical finance that large jump events of prices play a special role
and lead to modifications of the Black-Scholes theory
\cite{ref25}. These events are often created
by special circumstances of the market and are not explainable
by superstatistics. We repeated the calculation of $\tau$, $T$ and $|\epsilon|$ for data sets excluding the outliners. In practice we removed data points with $|u_i|>7$ (approximately $0.05\%$ of the data points). The bottom part of Table~\ref{tab} shows the newly obtained values for 
$\tau$, $T$, $|\epsilon|$ and the total kurtosis $\kappa_{1,n}$
when the outliners are removed. By comparing the values of the superstatistical parameters in Table~\ref{tab} one observes that $\tau$ and $T$ are hardly influenced by the outliners. However, notice the large decrease of the values of $\kappa_{1,n}$ and $|\epsilon|$, in particular
for the data sets with $\delta=1,2$. Removing approximately $0.05\%$ of the data points results in a decrease of the values of $\kappa_{1,n}$ and $|\epsilon|$ by a factor of the order $\sim5$ and $\sim100$ respectively. We conclude that both conditions $\tau/T<<1$ and $|\epsilon|<<1$ hold for the economical data sets with $\delta=1,2$ once the outliners are removed. Therefore, we will focus our analysis in the following part to these data sets. 

After the calculation of $T$ one can construct the distributions $p_T(u)$ and $f_T(\beta)$. Fig.~\ref{fig09} shows the original distributions $P(u)$ of the normalized log returns together with the result $p_T(u)$ of the superstatistical approach after the first approximation. There is an excellent agreement between $P(u)$ and $p_T(u)$ for all four cases. One also ends up with $N\approx15.000/T$ different values for the inverse variance $\beta_{T,l}$. We constructed a histogram with these values and tried to approximate this histogram with some well-known distributions, the gamma distribution (\ref{fbeta_surro12}), the lognormal distribution (\ref{poi}) and the inverse gamma distribution
\begin{eqnarray}\label{fbeta_infg}
f_T(\beta)&=&\frac{\theta^\alpha}{\Gamma(\alpha)}\beta^{-\alpha-1}e^{-\theta/\beta}.
\end{eqnarray}
These three distributions were motivated in \cite{ref02}. The resulting distributions are shown in Fig.~\ref{fig10}. We also constructed the distributions $p_{T,f}$, see Fig.~\ref{fig09}. By inspecting Fig.~\ref{fig09} and Fig.~\ref{fig10} we conclude that certainly the inverse gamma distribution is not a good candidate to represent $f_T(\beta)$. However, it is much harder to distinguish between the lognormal distribution and the gamma distribution. Both distributions are a reasonably good approximation for $f_T(\beta)$ and result in a good agreement between $p_{T,f}(u)$ and $p_T(u)$. We conclude that the data sets under study can be described equally well by gamma superstatistics and lognormal superstatistics.

We also studied the effect of a random shuffling of the data sets. A shuffling of the data keeps the kurtosis and the distribution $P(u)$ unchanged but destroys the correlations in the time series. As a consequence, we expect a considerable change in the values of $\tau$, $T$ and $|\epsilon|$. We applied the shuffling operation to the data of the DJI index and the S\&P 500 index for $\delta=1,2,4$, again with outliners removed. We shuffled every data set $500$ times and averaged the values of $T$ and $|\epsilon|$ over these different runs. The result of these calculations can be seen in Table~\ref{tab2}. As expected, the random shuffling operation causes a decrease of the value of $T$ and an increase of the value of $|\epsilon|$ (compare Table~\ref{tab} and Table~\ref{tab2}). These observations confirm the fact that the original economical time series is superstatistical whereas the shuffled one is not. Let us connect this with the results obtained in section \ref{subsection_ts}. We previously provided evidence that the distribution $P(u)$ of the economical data sets can be approximated very well by $p_{T,f}(u)$, see expression (\ref{Hu_2}) with $f_T(\beta)$ being a gamma distribution. In this case, the integral of expression (\ref{Hu_2}) can be evaluated analytically as
\begin{eqnarray}
p_{T,f}(u)&=&\frac{\Gamma\left(\alpha+\frac12\right)}{\Gamma(\alpha)}\sqrt\frac\theta\pi 2^\alpha\left(2+\theta u^2\right)^{-\alpha-1/2}.
\end{eqnarray}
This distribution is a q-Gaussian distribution, see expression (\ref{ts}), 
with 
\begin{eqnarray}
\alpha=\frac1{q-1}-\frac12&\textrm{and}&\theta=b(q-1).
\end{eqnarray}
In section \ref{subsection_ts} we studied the dependence of the value of $|\epsilon|$ as a function of the kurtosis for surrogate data sets containing q-Gaussian random variables. Because the economical data sets are in good approximation q-Gaussian distributed, it is no surprise that the values of the kurtosis ($\sim6$) and $|\epsilon|$ ($\sim0.3$) of Table~\ref{tab2} are of the same order of magnitude as the values that can be read off the curve shown in Fig.~\ref{fig05}. In other words, the randomly shuffled log returns can be modeled in good approximation with the q-Gaussian random variables studied in section \ref{subsection_ts}.

\begin{table}
\begin{tabular}{|c|c|ccc||ccc|}
\hline
&&S\&P&500&&DJI&&
\\\hline
$\delta$&&G&L&I&G&L&I
\\\hline
1&$\alpha$&2.02&0.78&1.74&2.05&0.84&1.28
\\
&$\theta$&1.05&0.56&2.45&0.96&0.51&1.64
\\
2&$\alpha$&2.04&0.80&1.54&2.10&0.77&1.47
\\
&$\theta$&0.98&0.49&2.01&0.88&0.44&1.83
\\\hline
\end{tabular}
\caption{\label{tab4} The values of the parameters $\alpha$ and $\theta$, used to construct the distributions shown in Fig.~\ref{fig09}, Fig.~\ref{fig10} and Fig.~\ref{fig11}. The letters G, L and I stand for the gamma distribution (\ref{fbeta_surro12}), the lognormal distribution (\ref{poi}) and the inverse gamma distribution (\ref{fbeta_infg}), respectively. }
\end{table}
\begin{table}
\begin{tabular}{|c|cccc||cccc|}
\hline
&S\&P&500&&&DJI&&&
\\\hline
$\delta$&$\tau$&$T$&$|\epsilon|$&$\kappa_{1,n}$&$\tau$&$T$&$|\epsilon|$&$\kappa_{1,n}$
\\\hline
1&$<$1&8&0.30&6.6&$<$1&9&0.32&6.3
\\
2&$<$1&9&0.30&6.1&$<$1&10&0.32&6.0
\\
4&$<$1&10&0.30&5.6&$<$1&11&0.30&5.5
\\\hline
\end{tabular}
\caption{\label{tab2} The values of the superstatistical parameters extracted from data series containing the randomly shuffled normalized log returns $u_i$ of the DJI index and the S\&P 500 index for three different values of $\delta$ (with largest values of $|u_i|$ removed).}
\end{table}
\begin{figure}
\begin{center}
\parbox{0.48\textwidth}{\includegraphics[width=0.48\textwidth]{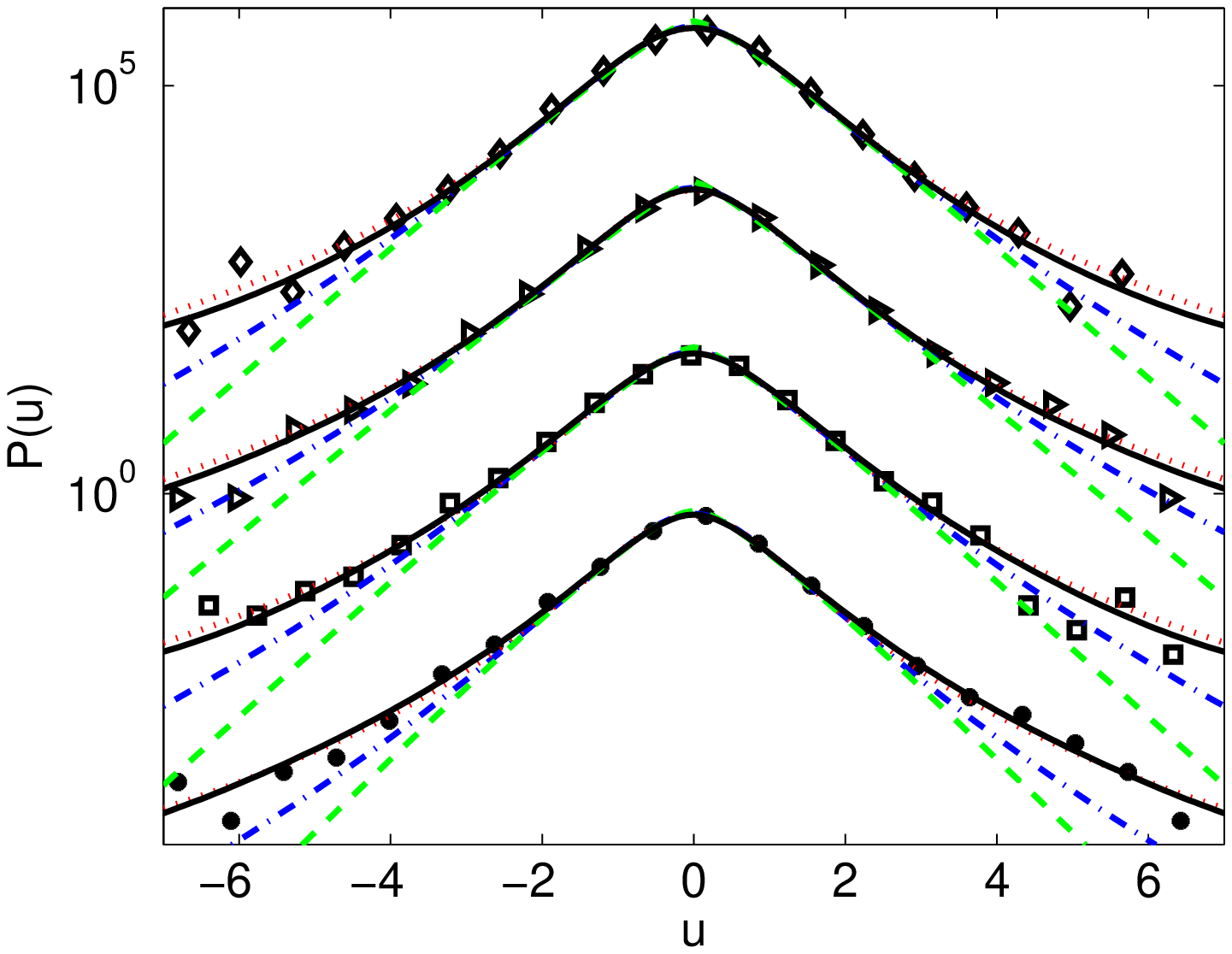}}
\hfill
\parbox{0.48\textwidth}{\includegraphics[width=0.48\textwidth]{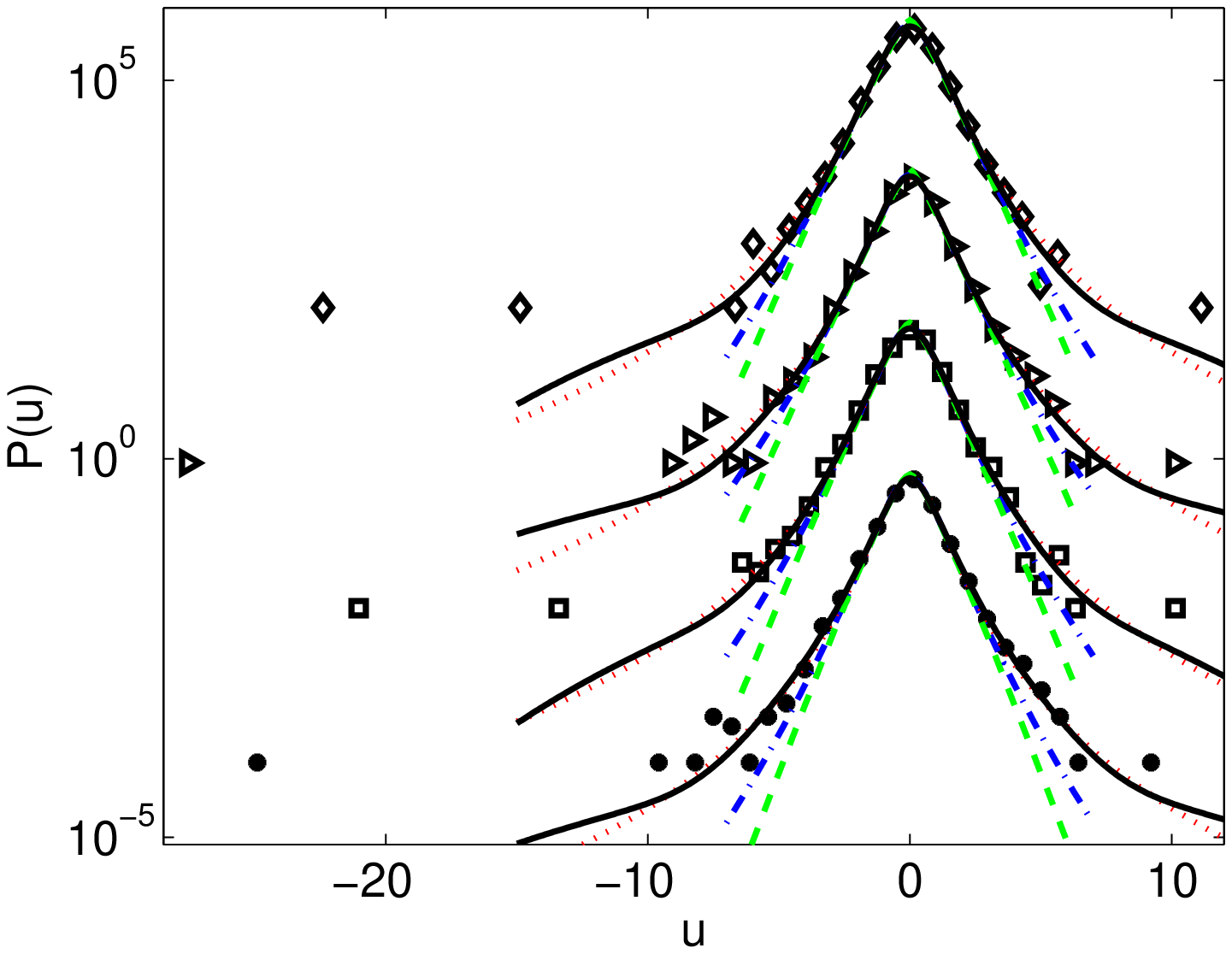}}
\caption{\label{fig11} (Colour online) Same picture as Fig.~\ref{fig09}
but with outliners included.}
\end{center}
\end{figure}
Finally, we studied how our superstatistical
techniques were influenced by extreme events (outliners) such as
stock market crashes.
We formally repeated our evaluation of the distribution $p_T(u)$ for the data series including the outliners, see Fig.~\ref{fig11}. This figure still shows excellent agreement between the superstatistical approximation $p_T(u)$ and the original distribution $P(u)$ for small and intermediate values of $u$ but fails
to reproduce the data in the region $|u|>10$.
The reason for that is quite clear: The basic assumption of superstatistics,
namely 'local' equilibrium, is violated during periods of 
very high market volatility. Again we approximated the extracted
$f_T(\beta)$ with the $3$ aforementioned distributions and construct the corresponding distributions $p_{T,f}(u)$.
When outliners are included, as Fig.~\ref{fig11} shows,
the best agreement between $p_T(u)$ and $p_{T,f}$ is obtained for
the gamma distribution $f_T(\beta)$. This analysis suggests that 
without eliminating the outliners gamma superstatistics is best suited 
to model realistic share-price dynamics.

\section{Alternative Approaches}\label{alt_rul}
\begin{figure}
\begin{center}
\parbox{0.48\textwidth}{\includegraphics[width=0.48\textwidth]{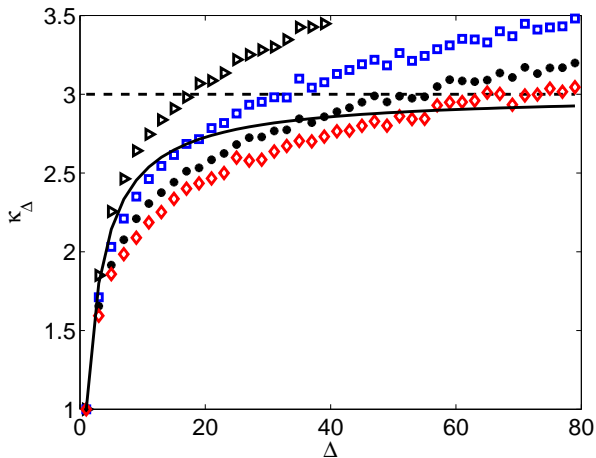}}
\caption{\label{fig12} (Colour online) Difference between the two
 definitions of the long time scale $T$ for time series containing 
the normalized log returns (with outliners removed) of the S\&P 500 index 
for $4$ different values of $\delta$. The average kurtosis 
$\kappa_\Delta$ of these time series is plotted
for $\delta=1$ ($\triangleright\triangleright\triangleright$), $\delta=2$ ({\tiny $\square\square\square$}), $\delta=3$ ($\bullet\bullet\bullet$), $\delta=4$ ($\diamond\diamond\diamond$). One definition fixes the value of $T$ by $\kappa_T=3$ (dashed line), while the other definition uses the equation $\kappa_T=3T/(2+T)$ (solid line).}
\end{center}
\end{figure}
The definition of a long superstatistical time scale $T$ is not unique
and alternative approaches can be considered in principle.
In this section we introduce an alternative definition for the long time scale 
and briefly repeat the study of the time series examined in sections 
\ref{section_dist_fb} and \ref{sect_eco}, thus testing for
the robustness of our methods.

Since superstatistics is based on a superposition of Gaussian distributions,
one might consider a suitable weighting of relevant observables with 
local Gaussian statistics.
An interesting alternative approach is the comparison of the average kurtosis $\kappa_\Delta$ of the time series at hand with the expectation of the kurtosis $\kappa(\Delta)$ of a time series with length $\Delta$ averaged over Gaussian statistics. In appendix \ref{appb} we show that $\kappa(\Delta)=3\Delta/(2+\Delta)$. The long time scale $T$ can then be extracted from the condition
\begin{eqnarray}\label{def_lon_alt}
\kappa_T&=&\kappa(T)=3\frac T{2+T}.
\end{eqnarray}
This definition for $T$ coincides with expression (\ref{def_lon}) in the 
limit $T\rightarrow\infty$. As a consequence, it is no surprise that 
for the turbulence data the resulting values of the long time scales almost coincide: $T=186$ ($\kappa_T=3$) and $T=170$ ($\kappa_T=\kappa(T)$).
We also checked that there is no visible change of the
distributions
$f_T(\beta)$ or $p_T(u)$ in Fig.~\ref{fig06} and Fig.~\ref{fig07}
with this new method. 

For the economical time series the typical values
of $T$ are smaller as compared to the
turbulence data, and hence stronger differences might arise
in this case.
Fig.~\ref{fig12} shows the average kurtosis $\kappa_\Delta$ for time series
containing the normalized log returns (with outliners removed) of the 
S\&P 500 index for different values of $\delta=1,2,3,4$. As expected, 
the obtained values of the long time scales are different for the two
methods, $T=/;17;36;52$ ($\kappa_T=\kappa(T)$) and $T=17;32;52;73$ ($\kappa_T=3$) for the time series with $\delta=1,2,3,4$ respectively. Here the slash
$/$ indicates that no solution is found, due to a lack
of intersection of the relevant curves in Fig.~12. Most importantly,
however, we checked that again
the corresponding distributions $f_T(\beta)$ or $p_T(u)$
 obtained for the new method 
are almost indistinguishable from Fig.~\ref{fig09} and 
Fig.~\ref{fig10}. This shows that our results are robust 
and that our main conclusions about the time series at hand do not
change.

In section \ref{mod_zelf} we derived a constraint $|\epsilon|<<1$ that 
should hold if a given time series is superstatistical.
For the alternative rule, expressions (\ref{def_theta}) and 
(\ref{rule_extra}) must be replaced by
\begin{eqnarray}
\theta_{T,l}&=&\langle u^4\rangle_{T,l}-3\frac T{2+T}\langle u^2\rangle_{T,l}^2
\cr 
\epsilon&=&\frac13\frac{2+T}T\frac{\sum_{j=1}^N\theta_{T,j}}{\sum_{j=1}^N\langle u^2\rangle_{T,j}^2}.
\end{eqnarray}
With this expression for $\epsilon$, we obtain
 $|\epsilon|=/;0.03;0.003;0.06$ ($\kappa_T=\kappa(T)$) for the time 
series with $\delta=1,2,3,4$ respectively, whereas
previously we had 
with expressions (\ref{def_theta}) and 
(\ref{rule_extra}) the values $|\epsilon|=0.02;0.01;0.03;0.05$ ($\kappa_T=3$). 
All these numbers are small, so our conclusions don't change and are robust,
though it is noticed that the new method apparently fails to provide
an $\epsilon$ and $T$-value
for the case $\delta =1$. Further alternative definitions can be given,
and the methods can be extended to include tests of
statistical significance.

\section{Discussion}
In this paper we have presented a general technique to study a given experimental time series with superstatistics. The crucial assumption of superstatistical modeling is that the dynamical description can be split into two levels that have a large time scale separation. Then, the complete data set 
can be divided into different time slices of length $T$ for which the variables are Gaussian distributed with a certain 
fixed inverse variance $\beta$. However, the value of $\beta$ varies from slice to slice. If the number of time slices is large enough one can construct a histogram with these values and approximate this histogram with a known distribution.We proposed a general procedure to introduce the different approximations of the superstatistical approach. Our method is as follows
\begin{itemize}
\item Extract the short $\tau$ time scale (\ref{kortetijd}), the long $T$ time scale (\ref{kazondergem},\ref{def_lon}) and the value of $\epsilon$ (\ref{rule_extra}) out of the time series at hand. 
\item Check whether the following two inequalities $\tau/T<<1$ and $|\epsilon|<<1$
are satisfied. Then construct the distribution $p_T(u)$ 
given by (\ref{Hu_1}) which is a first approximation for the original distribution $P(u)$.
\item Search for a good fit to the histogram of the values of $\beta$ with a known distribution $f(\beta)$. Then construct the distribution $p_{T,f}(u)$ (\ref{Hu_2}) which is a second approximation for $P(u)$. Only when a good fit is obtained together with $p_{T,f}(u)\approx p_T(u)$ one can conclude that 
the second approximation of the superstatistical modeling approach is valid.
\end{itemize}
We tested this method with several surrogate data sets and showed that 
our method is able to extract the correct information out of a given data set. We then applied the proposed techniques to two real experimental data series, velocity time series measured in turbulent Taylor-Couette flow and time series containing the normalized log returns of the closing prices of some stock market indices. For the turbulence data the inequalities $\tau/T<<1$ and $|\epsilon|<<1$ 
were immediately seen to hold, whereas for the share prices
the outliners had to be removed first in order to obtain a
realistic superstatistical model. We conclude that the superstatistical approach can be successfully used to study both data sets.

Since the early work of Black and Scholes \cite{ref26}, various
techniques borrowed from the field of the theoretical physics 
were successfully used to study the evolution of stock markets prices and their derivatives. Some recent work in the context of option pricing involves for example the use of continuous-time random walks \cite{ref26a,ref26b}, perturbation expansions around the Black-Scholes formula \cite{ref27} and the use of path integrals \cite{ref28}. Other authors focus on 
stochastic volatility and its extraction from a long sequence of
data \cite{ref10,ref29,ref30}. Common in all these papers is the observation that the volatility of the log returns $u_i$ of the stock market prices is a stochastic variable with certain distribution. However the type of this distribution is still under debate. Often the volatility is defined as the average of $|u_i|$ over a time slice with certain length $\Delta$. This results in $N=n/\Delta$ values for the volatility, with $n$ the total number of data points. Then one constructs a histogram with these $N$ values and searches for the best fit to this distribution 
using some known distribution. Usually, one examines different values of $\Delta$ and observes that the result of the fitting procedure is not crucially dependent on the arbitrary choice of $\Delta$. Notice the differences with the method presented in this paper. The length of our time slices coincides with long time scale $T$. Therefore, in the context of superstatistics, the value of $\Delta$ is fixed by the definition of $T$. We also use another measure for the volatility. Instead of studying the average of $|u_i|$ over a time slice we study the inverse of the average of $u_i^2$. The reason for this is that in the second step of our method we need the distribution of the latter variable to construct the distribution $p_{T,f}(u)$. Our careful analysis of the different approximations of the superstatistical approach shows that for the
hydrodynamic turbulence data our techniques can be applied
directly whereas for the economical time series it is better to
first remove the outliners. When outliners are removed,
our data sets can be described equally well by gamma superstatistics and lognormal superstatistics. When outliners are included, gamma superstatistics seems
to do the best job.

Our approach is inspired by the theory of statistical hypothesis testing. In this context, the null-hypothesis is the assumption that the data series at hand can be described by the superstatistical approach. Then one has to calculate the values for $\tau$, $T$ and $\epsilon$. One accepts the null-hypothesis when $T$
exists and $\tau/T<<1$ and $|\epsilon|<<1$. An interesting topic for future research is a statistical analysis of the threshold behavior  of $\tau/T$ and $|\epsilon|$. Also other conjectures to falsify the null-hypothesis can be examined. 
Generally this work will further help to understand the behavior
of complex systems with time scale separation, making direct contact with
experimental measurements.

\begin{acknowledgments}
We are very grateful to Harry Swinney for providing us with the measured
turbulence data.
E.V.d.S. gratefully acknowledges private discussions with Bart Partoens.
\end{acknowledgments}

\appendix
\section{\ }\label{appa}
The superstatistical long time scale $T$ is defined as the value of $\Delta$ for which $\kappa_\Delta$ equals $3$, see expression (\ref{kazondergem}) for the definition of $\kappa_\Delta$. Assume that we have found the value of the long time scale. This means that the complete time series is divided in $N=\lfloor n/T\rfloor$ time slices of length $T$. The superstatistical model assumption is valid when the variables within these time slices are in good approximation Gaussian distributed with a certain variance $\langle u^2\rangle_{T,l}$, with $l=1\ldots N$. We defined $\theta_{T,l}$ as the deviation of the fourth momentum from $3\langle u^2\rangle_{T,l}^2$ in the $l$th time slice of length $T$, see expression (\ref{def_theta}). Continue by dividing the complete time series in $\lfloor n/sT\rfloor$ time slices of length $\Delta=sT$ with $s=2,\ldots N$. We will show in this appendix that the value of $\kappa_{sT}$ can be expressed as a function of $\langle u^2\rangle_{T,l}$ and $\theta_{T,l}$ only. The contribution of $\theta_{T,l}$ to the value of $\kappa_{sT}$ is a measure for the deviations from the Gaussian distribution in the time slices of length $T$.

First notice that the moments within the $l$th time slice of length $sT$ can be written as a sum of the moments within the time slices of length $T$
\begin{eqnarray}\label{apa1}
\langle u^k\rangle_{sT,l}&=&\frac1{sT}\sum_{i=1+(l-1)sT}^{lsT}u_i^k
\cr 
&=&\frac1s\sum_{j=1}^s\langle u^k\rangle_{T,s(l-1)+j},
\end{eqnarray}
with $l=1,2,\ldots \lfloor n/sT\rfloor$. This gives for the fourth momentum and $\kappa_{sT,l}$ 
\begin{eqnarray}\label{apa2}
\langle u^4\rangle_{sT,l}&=&\frac1s\sum_{j=1}^s\langle u^4\rangle_{T,s(l-1)+j}
\cr 
&=&\frac1s\sum_{j=1}^s\left(\theta_{T,s(l-1)+j}+\frac3{\beta_{T,s(l-1)+j}^2}\right)
\cr 
\frac{\kappa_{sT,l}-3}3&=&\frac1{\langle u^2\rangle_{sT,l}^2}\left(B_{sT,l}+\Theta_{sT,l}\right)
\end{eqnarray}
where
\begin{eqnarray}\label{defBTHETA}
B_{sT,l}&=&\frac1s\sum_{j=1}^s\frac1{\beta_{T,s(l-1)+j}^2}-\left(\frac1s\sum_{j=1}^s\frac1{\beta_{T,s(l-1)+j}}\right)^2
\cr 
\Theta_{sT,l}&=&\frac1{3s}\sum_{j=1}^s\theta_{T,s(l-1)+j}.
\end{eqnarray}
To understand the meaning of $B_{sT,l}$ and $\Theta_{sT,l}$, we study an example. Assume $l=1$ which means that we look for the first time slice of length $sT$. This time slice contains the first $s$ time slices of length $T$, with corresponding values of the inverse variance $\beta_{T,1},\ldots,\beta_{T,s}$ respectively. For this special case, the expressions for $B_{sT,l}$ and $\Theta_{sT,l}$ simplify to
\begin{eqnarray}
B_{sT,1}&=&\frac1s\sum_{j=1}^s\frac1{\beta_{T,j}^2}-\left(\frac1s\sum_{j=1}^s\frac1{\beta_{T,j}}\right)^2
\cr 
\Theta_{sT,1}&=&\frac1{3s}\sum_{j=1}^s\theta_{T,j}.
\end{eqnarray}
$B_{sT,l}$ is the variance of $1/\beta_{T,j}$ calculated over $s$ values of this parameter. $\Theta_{sT,l}$ is the average of $\theta_{T,j}$ over $s$ values of this parameter. $B_{sT,l}$ vanishes when the fluctuations of $\beta_{T,j}$ are small. $\Theta_{sT,l}$ vanishes when the Gaussian approximation for the time slices of length $T$ holds. 

We continue by summing the expression $(\kappa_{sT,l}-3)/3$ over all the time slices $\lfloor n/sT\rfloor$. This results in an expression for $(\kappa_{sT}-3)/3$ 
\begin{eqnarray}\label{kappa_approx}
\frac{\kappa_{sT}-3}3&=&\frac1{\lfloor n/sT\rfloor}\sum_{l=1}^{\lfloor n/sT\rfloor}\frac1{\langle u^2\rangle_{sT,l}^2}\left(B_{sT,l}+\Theta_{sT,l}\right).
\nonumber\\ 
&&
\end{eqnarray}
Notice that this formula explains the difference in profoundness of the minimum between the two graphs of Fig.~\ref{fig01}. For both curves, the contribution of $\sum_l\Theta_{sT,l}$ is small compared to the contribution of $\sum_lB_{sT,l}$ and can be ignored (remember these graphs are obtained for surrogate data, where we imposed Gaussian distributed random variables). However, the values of $\sum_l B_{sT,l}$ are clearly different for the two graphs, because of the difference in the imposed distribution $f_S(\beta)$, see Fig.~\ref{fig02}. The narrower the distribution $f_S(\beta)$, the smaller the fluctuations of the parameter $\beta_{T,j}$, the less profound the minimum of the graph of $|\kappa_\Delta-3|/3$. 

The contribution of $\Theta_{sT,l}$ to $(\kappa_{sT,l}-3)/3$ vanishes when the Gaussian approximation in the time slices of length $T$ holds. So this term measures the deviations from pure Gaussian behavior in these time slices. We propose to study the difference between the exact value of $(\kappa_{sT}-3)/3$ with $s=2,3,\ldots$ and expression (\ref{kappa_approx}) with $\Theta_{sT,l}=0$ for all $l$ to decide whether a given time series can be described within the superstatistical approach. Clearly, this difference will show fluctuations as a function of $s$. Therefore, it is reasonable to evaluate this difference in the limit $s\rightarrow N=\lfloor n/T\rfloor$ in order to quantify the influence of $\Theta_{sT,l}$ to the exact value of $(\kappa_{sT}-3)/3$. For the special case $s=N$, formula (\ref{apa1}) simplifies to
\begin{eqnarray}
\langle u^k\rangle_{NT,1}&=&\frac1N\sum_{j=1}^N\langle u^k\rangle_{T,j}.
\end{eqnarray}
Then we obtain for $\kappa_{NT,1}$ (or $\kappa_{NT}$, because there is only one time slice for this special case)
\begin{eqnarray}\label{kappa_multiples}
\kappa_{NT}&=&\left(\frac1N\sum_{j=1}^N\langle u^2\rangle_{T,j}\right)^{-2}\frac1N\sum_{j=1}^N\left[3\langle u^2\rangle_{T,j}^2+\theta_{T,j}\right].
\nonumber\\
&&
\end{eqnarray}
Notice that for large data series, the kurtosis of the complete data set $\kappa_{n,1}$ is in good approximation equal to $\kappa_{NT}$ because $N=\lfloor n/T\rfloor\approx n/T$. When the Gaussian approximation is reasonable in the time slices of length $T$, the contribution of the term $\sum_j\theta_{T,j}$ to the value of $\kappa_{NT}$ will be small as compared to the contribution of the term $3\sum_j\langle u^2\rangle_{T,j}^2$. Therefore, we expect the Gaussian approximation to hold when $|\epsilon|<<1$ with
\begin{eqnarray}
\epsilon&=&\frac13\frac{\sum_{j=1}^N\theta_{T,j}}{\sum_{j=1}^N\langle u^2\rangle_{T,j}^2}.
\end{eqnarray}
The value of $\epsilon$ basically measures the contribution of the deviations from the Gaussian approximation in the time slices of length $T$ to the value of the kurtosis of the complete data set. 

\begin{figure}
\begin{center}
\parbox{0.48\textwidth}{\includegraphics[width=0.48\textwidth]{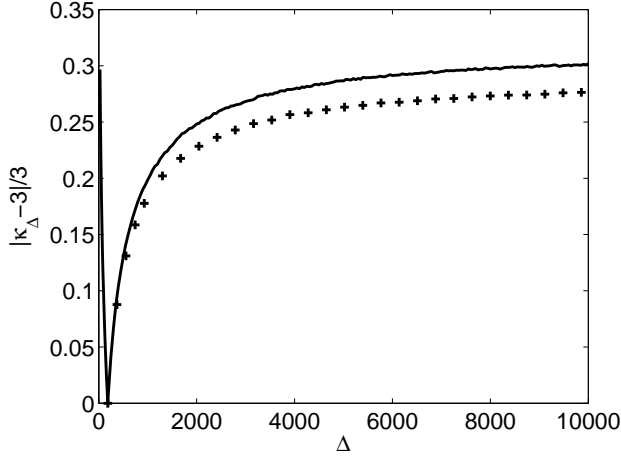}}
\caption{\label{fig13} The solid line shows the value of $|\kappa_\Delta-3|/3$ calculated with expression (\ref{kazondergem}). The evaluation of expression (\ref{kappa_approx}) with $\Theta_{sT,l}=0$ for all $l$ is represented by ({\scriptsize$+++$}). The difference between the solid line and the curve represented by ({\scriptsize$+++$}) is a consequence of the deviations from the Gaussian approximation in the time slices of length $T$.}
\end{center}
\end{figure}

Finally we illustrate the physical interpretation of the formulas we derived above by applying them to the real experimental data that we already
studied in section \ref{section_dist_fb}. We repeated the calculation of $|\kappa_\Delta-3|/3$ as a function of $\Delta$ with formula (\ref{kazondergem}). The result can be seen in Fig.~\ref{fig13} (solid line). Then, we evaluated expression (\ref{kappa_approx}) with $\Theta_{sT,l}=0$ for all $l$ ($+++$). Clearly, the difference between the solid line and the curve represented by ($+++$) is small. This shows that the contribution of the term $\sum_l\Theta_{T,j}/\langle u^2\rangle^2_{T,j}$ to the value of $\kappa_{sT}$ is small as compared to the contribution of the term $\sum_jB_{T,j}/\langle u^2\rangle^2_{T,j}$. Hence the Gaussian approximation in the time slices of length $T$ is reasonable because the term $\sum_l\Theta_{T,j}/\langle u^2\rangle^2_{T,j}$ measures the deviation from this approximation. The value of $|\epsilon|$ corresponds to the difference between the solid line and the curve represented by ($+++$) in the limit $s\rightarrow N$. For this example we obtain $|\epsilon|=0.0205<<1$. 

\section{\ }\label{appb}
In this appendix we calculate the mathematical expectation of the kurtosis $\kappa(\Delta)$ of a time series with length $\Delta$ averaged over a Gaussian statistics
\begin{eqnarray}
\kappa(\Delta)&=&\Delta\left(\frac\beta{2\pi}\right)^{\frac \Delta2}\int_{-\infty}^{+\infty}dx_1\ldots\int_{-\infty}^{+\infty}dx_\Delta\times
\cr
&&\frac{\sum_{i=1}^\Delta x_i^4}{\left(\sum_{i=1}^\Delta x_i^2\right)^2}\exp\left({-\frac12\beta\sum_{i=1}^\Delta x_i^2}\right).
\end{eqnarray}
To calculate these $\Delta$ integrals it is convenient to consider a transformation to hyperspherical coordinates \cite{ref31}
\begin{eqnarray}
\left\{
\begin{array}{ccl}
x_1&=&r\cos\phi_1
\cr 
x_2&=&r\sin\phi_1\cos\phi_2
\cr 
&\vdots&
\cr
x_{\Delta-1}&=&r\sin\phi_1\ldots\sin\phi_{\Delta-2}\cos\phi_{\Delta-1}
\cr 
x_\Delta&=&r\sin\phi_1\ldots\sin\phi_{\Delta-2}\sin\phi_{\Delta-1}
\end{array}
\right.
\end{eqnarray}
with $r\in[0..\infty]$, $\phi_1,\ldots,\phi_{\Delta-2}\in[0..\pi]$ and $\phi_{\Delta-1}\in[0..2\pi]$. The Jacobian of this transformation is
\begin{eqnarray}
r^{\Delta-1}\sin(\phi_1)^{\Delta-2}\sin(\phi_2)^{\Delta-3}\ldots\sin(\phi_{\Delta-2}).
\end{eqnarray}
With the use of the well-known formula for the volume of a $\Delta$-dimensional sphere with radius $R$
\begin{eqnarray}
\frac{\pi^{\Delta/2}}{\Gamma(\Delta/2+1)}R^\Delta,
\end{eqnarray}
one can easily check that the expression for $\kappa(\Delta)$ simplifies to
\begin{eqnarray}
\kappa(\Delta)&=&3\frac\Delta{2+\Delta}.
\end{eqnarray}

\end{document}